\documentclass[pra,twocolumn]{revtex4-1}
\usepackage{hyperref} 

\usepackage{graphicx} 
\usepackage{subfig}
\usepackage{color}
\usepackage{filecontents}
\usepackage{soul}
\usepackage{mathrsfs}
\usepackage{tikz,bm} 
\usepackage{circuitikz}
\usepackage{tikz}
\usepackage{physics}
\usepackage{mathtools}
\usepackage[T1]{fontenc}
\usepackage[outline]{contour}
\usepackage{relsize}

\usetikzlibrary{shapes.geometric}
\usetikzlibrary{decorations.pathmorphing}
\usetikzlibrary{shapes.arrows, fadings}
\usetikzlibrary{arrows,snakes,shapes}

\definecolor{mycolor}{rgb}{1,0.2,0.3}
\definecolor{brightgreen}{rgb}{1.0, 1.0, 1.0}
\definecolor{britishracinggreen}{rgb}{0.0, 0.26, 0.15}
\definecolor{cadmiumgreen}{rgb}{0.0, 0.42, 0.24}
\definecolor{ceruleanblue}{rgb}{0.16, 0.32, 0.75}
\definecolor{darkelectricblue}{rgb}{0.33, 0.41, 0.47}
\definecolor{darkpowderblue}{rgb}{0.0, 0.2, 0.6}
\definecolor{dt}{rgb}{1.0, 0.66, 0.07} 
\definecolor{emerald}{rgb}{0.31, 0.78, 0.47}
\definecolor{palatinatepurple}{rgb}{0.41, 0.16, 0.38}
\definecolor{pastelviolet}{rgb}{0.8, 0.6, 0.79}
\definecolor{br}{rgb}{0.5, 0.05, 0.01}
\definecolor{chosen_color}{RGB}{3, 207, 252}

\pgfdeclareverticalshading{pgf@lib@fade@north}{100bp}
{color(0bp)=(pgftransparent!0);
 color(30bp)=(pgftransparent!0);
 color(50bp)=(pgftransparent!80); 
 color(65bp)=(pgftransparent!100);
 color(100bp)=(pgftransparent!100)}%
\pgfdeclarefading{exponentialtop}{%
\pgfuseshading{pgf@lib@fade@north}%
}

\pgfdeclareverticalshading{pgf@lib@fade@north}{100bp}
{color(0bp)=(pgftransparent!100);
 color(35bp)=(pgftransparent!100);
 color(50bp)=(pgftransparent!80); 
 color(70bp)=(pgftransparent!00);
 color(100bp)=(pgftransparent!00)}%
\pgfdeclarefading{exponentialbottom}{%
  \pgfuseshading{pgf@lib@fade@north}%
}

\newcommand{\be}{\begin{equation}}
\newcommand{\ee}{\end{equation}}
\newcommand{\bea}{\begin{eqnarray}}
\newcommand{\eea}{\end{eqnarray}}
\newcommand*{\myeqref}[2][Eq.~]{%
\hyperref[{#2}]{#1(\ref*{#2})}%
}
\def\equationautorefname#1#2\null{%
Eq.#1(#2\null)%
}

\usepackage{dcolumn}
\usepackage{bm}
\usepackage{amssymb,amsmath}
\usepackage{color}
\usepackage{float}
\usepackage{tikz}
\usetikzlibrary{arrows}

\definecolor{DarkGreen}{rgb}{0,0.6,0.2}

\usepackage{mathrsfs}

\begin{document}
\title{Operational boundaries, threshold conditions, and self-quenching in single-emitter microcavity lasers}
\author{Alexandra Gospodinov$^{1}$} 
\author{Celia Powers$^{1}$}  
\author{Imran M. Mirza$^{1,2}$}
\email{mirzaim@miamioh.edu}
\affiliation{$^{1}$Macklin Quantum Information Sciences,Department of Physics, Miami University, Oxford, OH 45056, USA\\
$^{2}$Department of Computer Science \& Software Engineering, Miami University, Oxford, OH 45056, USA}
\date{\today}

\begin{abstract}
We theoretically examine the emission dynamics of a three-level cascade single-emitter microcavity laser using the open-system Lindblad master equation. In our numerical simulations, we expand the Hilbert space dimensions to obtain converged observables free from truncation artifacts. We notice that the intracavity field builds up continuously to a peak value of $\langle n_p \rangle_{\rm max}$, bypassing the sharp thermodynamic thresholds of conventional ensemble lasers. By tracking individual radiative rates and for the selected set of parameters, we isolate a physical threshold at a normalized pump rate of $\Gamma/\gamma_{12} \approx 2.45$, where net stimulated emission overcomes spontaneous noise. Crucially, comparing these numerical master equation solutions against factorized semiclassical models reveals that mean-field approximations omit a vital light-matter correlation drag force, leading to a substantial overestimation of the peak field intensity by more than 10\%. We notice that the transient Wigner function undergoes complete phase diffusion into a radially symmetric ring due to the absence of a phase-locking ensemble. Furthermore, steady-state noise calculations confirm that discrete incoherent pumping fluctuations restrict the field to strictly super-Poissonian statistics (with the Mandel parameter $Q > 0$ and the second-order coherence function with zero delay $g^{(2)}(0) > 1$), establishing the classical shot-noise limit as a strict lower bound. At high pumping rates, population trapping in the upper level creates an excitation bottleneck that suppresses the cavity field and causes self-quenching. Our results provide an essential baseline for optimizing solid-state quantum dot and color-center nanophotonic architectures for lasing as well as for the generation of non-classical states of light.
\end{abstract}


\maketitle

\section{\label{sec:I} Introduction}
Single atoms or quantum emitters coupled to microscopic optical cavities push light-matter interfaces to their fundamental quantum limits, providing a highly controllable architecture for examining open quantum dynamics while establishing strong coupling regime at the level of single atoms and single photons \cite{mu1992one, mckeever2003experimental, thon1992observation}. While conventional macroscopic lasers rely on collective atomic ensembles as a gain medium to overcome dissipative losses \cite{scully1997quantum}, single-emitter cavity quantum electrodynamics (cQED) setups leverage Purcell enhancement to direct spontaneous emission efficiently into a singular resonant optical mode \cite{walther2006cavity, rice1994photon}. This mechanism enables continuous, smooth intensity growth and amplitude-stabilized operation that can approach the classical shot-noise limit \cite{ginzel1993quantum, rempe1990observation}. Over the past few decades, these single-atom systems have progressed from fundamental testbeds for cQED into essential building blocks for engineering scalable nanolasers, single-photon sources, and distributed nodes for quantum communication networks \cite{kimble2008quantum, an1994microlaser, stolk2024metropolitan, wang2025scalable}.

Historically, single-emitter lasing theory began with foundational models mapping high spontaneous emission fractions to the erasure of traditional thermodynamic thresholds \cite{mu1992one, fleischhauer1994relation, rice1994photon}. Lasing operation was subsequently validated in different experimental setups, including trapped Cesium atoms \cite{mckeever2003experimental}, single ions \cite{dubin2010quantum}, and superconducting artificial charge qubits \cite{astafiev2007single}. Parallel developments in semiconductor nanophotonics successfully demonstrated thresholdless scaling loops using self-assembled quantum dots embedded in micropillars and photonic crystal defects \cite{ritter2010emission, gartner2011two, yamamoto1991coherence, reitzenstein2008lasing}. Subsequent theoretical models explored alternative configurations to optimize these nanoscale sources, investigating sub-fractional photon boundaries \cite{boozer2008laserlike}, Quantum Zeno thermal collapse limits \cite{ashhab2009single}, closed-form linewidth dependencies \cite{larionov2013quantum}, Andreev bound-state loops \cite{rastelli2019single}, and inversionless gain cycles using electromagnetically induced transparency \cite{villas2020continuous}. Most recently, research has shifted toward multi-level dot structures to enable non-degenerate multi-photon lasing and continuous-variable entanglement channels \cite{hazra2024nondegenerate, addepalli2024cooperative}.

\begin{figure*}
\begin{center}
\includegraphics[width=4.2in, height=1.8in]{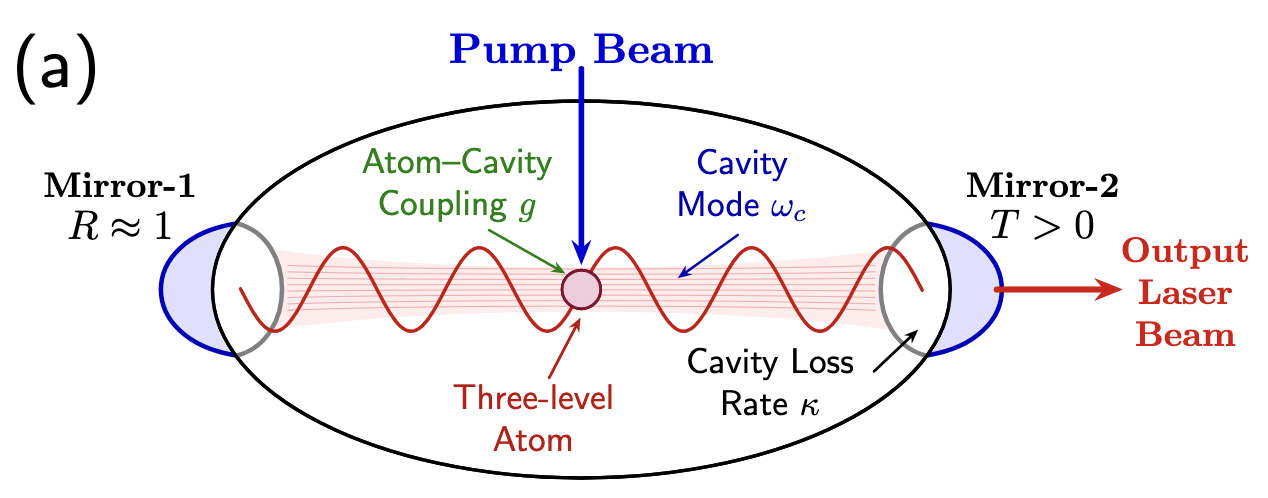} 
\includegraphics[width=2.35in, height=1.85in]{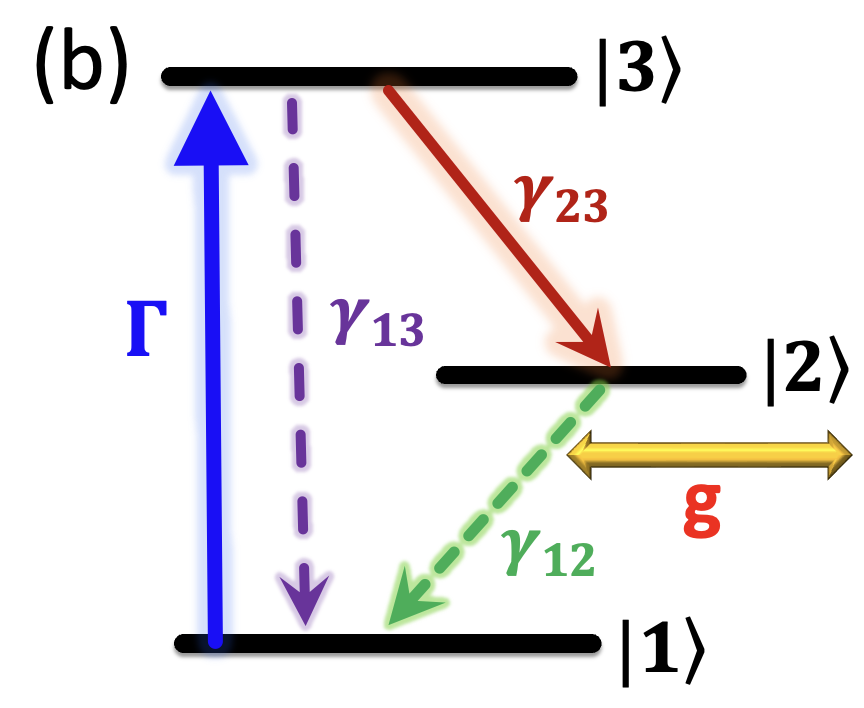} 
\captionsetup{
format=plain,
margin=1em,
justification=raggedright,
singlelinecheck=false
}
\caption{(Color online) (a) Physical setup under study, consisting of a single three-level emitter trapped inside an optical cavity. The resonator is bounded by a highly reflective back mirror (Mirror-1, $R \approx 1$) and a partially transparent out-coupling mirror (Mirror-2, $T > 0$). Photons leak out of the resonator at a characteristic cavity loss rate $\kappa$ to form the output laser beam. (b) Energy-level configuration of our three-level quantum emitter. We examine a purely incoherent pumping mechanism, where an external drive populates the highly excited state $|3\rangle$ from the ground state $|1\rangle$ at a rate $\Gamma$. The underlying dissipative relaxation pathways are governed by spontaneous radiative decay channels represented by the rates $\gamma_{13}$, $\gamma_{23}$, and $\gamma_{12}$. The active $|2\rangle \leftrightarrow |1\rangle$ optical transition is strongly coupled to the single resonant cavity mode with an coupling parameter $g$.
}\label{Fig1}
\end{center}
\end{figure*}

While these historical works established the essential foundations for mapping single-emitter cQED architectures into lasing regimes, early theoretical treatments were frequently constrained by semiclassical mean-field factorizations or highly restricted computational basis cutoffs. In this paper, we study the steady-state and transient emission dynamics of an incoherently pumped three-level single-emitter microcavity laser by solving the full open-system Lindblad master equation. We expand the computational Hilbert space dimensions to a systematically verified ceiling ($\mathcal{N}=60$) where the top-state population satisfies $P(\mathcal{N}-1) < 10^{-7}$. This approach allows us to extract truncation-invariant observables that represent physical states completely free from numerical clipping artifacts. The primary novelty of this work lies in identifying and quantifying the exact light-matter correlation drag force that is dropped when applying standard semiclassical mean-field decouplings. We examine the precise boundaries where factorized mean-field models fail under a given experimentally feasible set of parameters. We notice that the semiclassical approach omits these multi-operator cross-correlations entirely, which creates a non-physical artifact that overestimates the peak cavity field intensity by more than 10\%. By evaluating individual radiative decay rates without approximation, we isolate a physical operational threshold at a normalized pump rate of $\Gamma/\gamma_{12} \approx 2.45$, where the net stimulated emission rate successfully overcomes the spontaneous noise floor. 

Furthermore, we evaluate the statistical properties of the generated field across the full pumping spectrum under realistic experimental constraints, mapping our parameters onto achievable solid-state quantum dot and color-center configurations. We study the transient time-evolution of the cavity field by computing its phase-space Wigner quasiprobability distribution. We notice that the distribution undergoes complete phase diffusion over time, broadening out into a radially symmetric donut ring. This absolute symmetry serves as a direct physical benchmark of the single-emitter architecture; without a macroscopic atomic ensemble to break the $U(1)$ gauge symmetry, the random phase kicks from the incoherent pump diffuse the field uniformly. Finally, our steady-state noise calculations show that discrete incoherent pumping fluctuations restrict the intensity profile to strictly super-Poissonian statistics (the Mandel parameter $Q>0,$ and the zero-delay second-order correlation function $g^{(2)}(0)>1$), proving that the classical shot-noise limit forms an unyielding lower bound for this single-atom system before high-pump population trapping triggers the phenomenon of self-quenching.

The rest of the paper is organized as follows. In Sec.~\ref{sec:II}, we present the theoretical framework, detailing the model Hamiltonian, dissipative description through the Lindblad master equation, and our numerical stability validation protocols. In Sec.~\ref{sec:III}, we present and discuss our results, focusing on the operational threshold criteria, multi-photon correlations, non-classical statistics, and a physical assessment of experimental feasibility. Finally, in Sec.~\ref{sec:IV}, we close with a summary of our main conclusions.

\section{\label{sec:II} Theoretical Model, Dissipative Dynamics, and Numerical Stability}
As illustrated in the schematic layout of Fig.~\ref{Fig1}(a), the physical configuration under study consists of a single three-level quantum emitter or atom trapped inside a high-finesse optical Fabry-Pérot cavity. The resonator structure of the cavity is bounded by a highly reflective back mirror ($\text{Mirror-1}, R \approx 1$) and a partially transparent out-coupling mirror ($\text{Mirror-2}, T > 0$). Photons leak out of the resonator at a characteristic cavity loss rate $\kappa$ to form the output laser beam. The inner electronic energy-level structure of the atomic emitter is shown in Fig.~\ref{Fig1}(b), defining a cascaded configuration with a ground state $|1\rangle$, an intermediate lasing state $|2\rangle$, and a topmost excited state $|3\rangle$. An external incoherent pump field with an pumping rate $\Gamma$ drives the transitions from $|1\rangle \to |3\rangle$. The active dipole transition transition channel $|2\rangle \leftrightarrow |1\rangle$ is strongly coupled to the single isolated resonant cavity mode at frequency $\omega_c$ with an atom-cavity coupling parameter $g$. The inner electronic structure of our multi-level atom is modeled through pseudo-Fermionic operators formally defined as:
\begin{equation}
    \hat{\sigma}_{ij} = |i\rangle\langle j|, \quad \text{for } i,j \in \{1, 2, 3\}.
\label{atomic_projectors}
\end{equation}
which obey non-canonical anti-commutation relations given by:
\begin{equation}
    \{\hat{\sigma}_{ij}, \hat{\sigma}_{k\ell}\} = \hat{\sigma}_{ij}\hat{\sigma}_{k\ell} + \hat{\sigma}_{k\ell}\hat{\sigma}_{ij} = \hat{\sigma}_{i\ell}\delta_{jk} + \hat{\sigma}_{kj}\delta_{i\ell},
\label{atomic_anticommutation}
\end{equation}
where $\delta_{jk}$ denotes the standard Kronecker delta index. The cavity mode is governed by bosonic creation and annihilation operators, $\hat{a}^\dagger$ and $\hat{a}$, which satisfy the commutation relation $[\hat{a}, \hat{a}^\dagger] = 1$. Under the dipole and rotating-wave approximations, the coherent dynamics of the atom-cavity coupled system are governed by the total system Hamiltonian ($\hbar = 1$):
\begin{align}
    \hat{\mathcal{H}} & = \hat{\mathcal{H}}_{\rm atom} + \hat{\mathcal{H}}_{\rm cav} + \hat{\mathcal{H}}_{\rm int}\nonumber\\ &= \sum_{j=2}^{3} \omega_{j1}\hat{\sigma}_{jj} + \omega_c \hat{a}^\dagger\hat{a} + ig \left(\hat{a}^\dagger\hat{\sigma}_{12} - \hat{a}\hat{\sigma}_{21}\right),
\label{full_hamiltonian}
\end{align}
where $\omega_{j1}$ denotes the transition frequency from the ground state to electronic level $|j\rangle$. The first two terms describe the free energies of the quantum emitter/atom and the cavity field, respectively. The final term represents the coherent light-matter interaction coupling between the $|2\rangle \leftrightarrow |1\rangle$ atomic transition to the cavity mode through the standard Jaynes-Cummings model \cite{jaynes1963comparison}. Note that we have separated out the factor of $i$ from the dipole matrix element to preserve real and symmetric coupling terms throughout subsequent calculations \cite{mu1992one,rice1994photon}.
\subsection{\label{subsec:master_equation} Open Quantum System Treatment}
To study the statistical transitions, population inversions, and dissipative dynamics applicable to a realistic cQED platform, we evaluate the reduced atom-cavity system density matrix $\hat{\rho}$ which obeys the following Lindblad master equation:
\begin{align}
    \frac{d\hat{\rho}}{dt} & = -i[\hat{\mathcal{H}}, \hat{\rho}] + \kappa \mathcal{D}[\hat{a}]\hat{\rho} + \gamma_{12} \mathcal{D}[\hat{\sigma}_{12}]\hat{\rho} + \gamma_{23} \mathcal{D}[\hat{\sigma}_{23}]\hat{\rho}\nonumber\\
    &+ \Gamma \mathcal{D}[\hat{\sigma}_{31}]\hat{\rho} + \sum_{k=2}^{3} \gamma_{d,k} \mathcal{D}[\hat{\sigma}_{kk}]\hat{\rho},
\label{lindblad_master}
\end{align}
where the standard dissipative superoperator acting on an arbitrary operator $\hat{O}$ is defined as:
\begin{equation}
    \mathcal{D}[\hat{O}]\hat{\rho} = 2\hat{O}\hat{\rho}\hat{O}^\dagger - \hat{O}^\dagger\hat{O}\hat{\rho} - \hat{\rho}\hat{O}^\dagger\hat{O}.
\label{dissipative_superoperator}
\end{equation}
The relaxation parameters represent the cavity field mirror transmission loss rate $\kappa$, the spontaneous decay rates ($\gamma_{12}$ and $\gamma_{23}$), and the continuous incoherent optical pumping drive ($\Gamma$) that populates level $|3\rangle$ directly out of the ground state $|1\rangle$, closing the excitation loop. The final summation introduces the pure dephasing rates ($\gamma_{d,2}$ and $\gamma_{d,3}$) governing the intermediate and upper electronic states, accounting for the localized phonon-scattering fluctuations characteristic of some of the cQED platforms (such as the solid-state environments).

\subsection{\label{subsec:numerical_stability} Tensor Product Structure and Numerical Stability}

To proceed with the numerical calculations, we note that the composite Hilbert space of the coupled system $\mathscr{H}$ is constructed as the tensor product of the constituent subsystems, $\mathscr{H} = \mathscr{H}_{\text{cav}} \otimes \mathscr{H}_{\text{atom}}$, where $\text{dim}(\mathscr{H}_{\text{atom}}) = 3$ and $\text{dim}(\mathscr{H}_{\text{cav}}) = \mathcal{N}$. To evaluate both the steady-state behavior and the transient dynamics of our system, we implement numerical simulations using the Quantum Toolbox in Python (QuTiP) \cite{lambert2026qutip}. Within this computational framework, we construct the full-dimension operators via explicit tensor embeddings with identity operators ($\hat{\mathbb{I}}$) according to:
\begin{align}
    \hat{a} &= \hat{a}_{\mathcal{N}} \otimes \hat{\mathbb{I}}_3, \\
    \hat{\sigma}_{ij} &= \hat{\mathbb{I}}_{\mathcal{N}} \otimes |i\rangle\langle j|,
\end{align}
where $\hat{a}_{\mathcal{N}}$ denotes the standard $\mathcal{N} \times \mathcal{N}$ bosonic annihilation matrix truncated in the Fock basis. 

To find the steady-state density operator $\hat{\rho}_{\text{ss}}$, we impose the stationary condition $d\hat{\rho}/dt = 0$, which maps the time-dependent Lindblad master equation (Eq.~\eqref{lindblad_master}) onto a static matrix equation. We execute this computation by vectorizing the $3\mathcal{N} \times 3\mathcal{N}$ density matrix into a column vector of dimension $(3\mathcal{N})^2$, represented in the standard double-ket notation as $\rvert\hat{\rho}_{\text{ss}}\rangle\rangle$. This algebraic transformation allows us to project the combined left- and right-acting coherent and dissipative superoperators into a single, unified Liouvillian matrix $\mathcal{L}$ acting in Liouville space, yielding:
\begin{equation}
    \mathcal{L} \rvert\hat{\rho}_{\text{ss}}\rangle\rangle = 0.
\label{liouvillian_flattened}
\end{equation}
Because the Liouvillian matrix $\mathcal{L}$ is highly sparse, we can isolate its null space efficiently. We utilize specialized sparse-matrix solvers within QuTiP to find the stationary solution vector of Eq.~\eqref{liouvillian_flattened}, subject to the unit-trace normalization condition $\text{Tr}\{\hat{\rho}_{\text{ss}}\} = 1$. 

Beyond the steady-state regime, we investigate the full transient dynamics by directly solving the time-dependent Lindblad master equation. Starting from a specified initial state $\hat{\rho}(0)$, we utilize QuTiP's master equation solver to numerically integrate the system of coupled differential equations up to a target time $t$. The solver dynamically propagates the density matrix $\hat{\rho}(t)$, preserving its Hermiticity and unit trace at each intermediate time step. From the full time-evolved density matrix $\hat{\rho}(t)$, we isolate the cavity field properties by performing a partial trace over the atomic degrees of freedom, yielding the time-dependent reduced density matrix $\hat{\rho}_{\text{field}}(t) = \text{Tr}_{\text{atom}}\{\hat{\rho}(t)\}$. This time-resolved state allows us to compute and reconstruct the phase-space evolution of the cavity field via the time-dependent Wigner function $W(\alpha, t)$, as reported in the next section. 

To ensure both our steady-state and time-resolved light-matter dynamics remain free from truncation errors, numerical clipping, or non-physical artifacts, we systematically verify our observables by expanding the computational basis cutoff from $\mathcal{N} = 40$ to $\mathcal{N} = 60$. We formally monitor this numerical stability across all simulation durations by tracking the population probability of the highest accessible Fock state within the cavity field:
\begin{equation}
    P(\mathcal{N}-1, t) = \langle \mathcal{N}-1 \rvert \hat{\rho}_{\text{field}}(t) \lvert \mathcal{N}-1 \rangle.
\label{highest_fock_pop}
\end{equation}
We consider a solution globally converged and independent of the basis size only when $P(\mathcal{N}-1, t) < 10^{-7}$ holds true for all time steps. Through this strict threshold we verified that the truncation boundary is securely positioned in the asymptotic tail of the photon distribution, confirming that our extracted threshold points, statistical trends, and phase-space results represent invariant physical states rather than constraints of a restricted numerical subspace.

\section{\label{sec:III} Results and Discussion}
In this section, we numerically solve the Lindblad master equation and the semiclassical rate equations under the aforementioned convergence conditions to analyze how the mean photon number $\langle n_p \rangle$, population inversion $\Delta_{12}$, stimulated emission rate $R_{\text{st}}$, and spontaneous emission rate $R_{\text{sp}}$ depend on the incoherent pumping rate $\Gamma/\gamma_{12}$. Since single-emitter lasers lack a sharp thermodynamic threshold, defining the exact onset of lasing requires comparing different criteria. Therefore, we analyze the system over a wide range of pumping rates to compare mainly three regimes of interest defined as: 
\begin{itemize}
    \itemsep=0pt \parsep=0pt \topsep=0pt \partopsep=0pt
    \item \textit{Sub-Threshold Weak Driving Regime:} The weak-excitation limit ($\Gamma/\gamma_{12} < 2.45$), where cavity dynamics are dominated by spontaneous emission noise and the stimulated gain cannot overcome cavity losses.
    \item \textit{Optimal Lasing Coherence Domain:} The intermediate region ($2.45 \le \Gamma/\gamma_{12} \le 25$), where a coherent photon field forms and the state statistics are governed by quantum fluctuations and system nonlinearities.
    \item \textit{High-Pump Self-Quenching Envelope:} The high-pump limit ($\Gamma/\gamma_{12} > 25$), where upper-level population trapping occurs, which we analyze using a semiclassical treatment.
\end{itemize}
In the following, we evaluate these threshold criteria and discuss the physical characteristics of each regime.

\subsection{\label{subsec:steady_state_profiles} Steady-State Behavior and Lasing Thresholds}
We calculate the steady-state properties of the single-emitter microcavity field by varying the normalized incoherent pump rate $\Gamma/\gamma_{12}$. To avoid numerical truncation artifacts, all observables are computed under verified convergence conditions. We obtain these values from the diagonal components of the stationary density matrix $\hat{\rho}_{\text{ss}}$ expressed in the composite Fock-atomic basis:
\begin{align}
    \rho_{n,i} \equiv \langle n, i \rvert \hat{\rho}_{\text{ss}} \lvert n, i \rangle,
\label{basis_matrix_elements}
\end{align}
where $n \in \{0, 1, \dots, \mathcal{N}-1\}$ is the cavity photon number and the index $i \in \{1, 2, 3\}$ represents the electronic states of the three-level cascade emitter. Using these definitions, we determine the mean intracavity photon number $\langle n_p \rangle$, the population inversion $\Delta_{12}$ on the $|2\rangle \to |1\rangle$ transition, and the corresponding radiative rates which are defined as:
\allowdisplaybreaks
\begin{subequations}
\begin{align}
    \langle n_p \rangle &= \text{Tr}\{\hat{a}^\dagger \hat{a} \hat{\rho}_{\text{ss}}\} = \sum_{n=0}^{\mathcal{N}-1} \sum_{i=1}^{3} n \rho_{n,i}, \label{np_trace_eq} \\
    \Delta_{12} &= \text{Tr}\{(\hat{\sigma}_{22} - \hat{\sigma}_{11})\hat{\rho}_{\text{ss}}\} = \sum_{n=0}^{\mathcal{N}-1} \left( \rho_{n,2} - \rho_{n,1} \right), \label{delta_trace_eq} \\
    G_{n} &= \frac{4g^2}{\gamma_{12} + \Gamma + 2\gamma_{d,2} + 2\kappa \left( n - \frac{1}{2} - \sqrt{n(n+1)} \right)}, \label{gn_coupling_eq} \\
    R_{\text{st}} &= \sum_{n=0}^{\mathcal{N}-1} G_{n} \bigg[ n \rho_{n,2} - (n + 1) \rho_{n+1,1} \bigg], \label{rst_trace_eq} \\
    R_{\text{sp}} &= \sum_{n=0}^{\mathcal{N}-1} G_{n} \rho_{n,2}. \label{rsp_trace_eq}
\end{align}\label{quantum_rate_system}
\end{subequations}
We note in Eq.~\eqref{gn_coupling_eq} a nonlinear light-matter coupling coefficient $G_n$ has been defined, which includes the line broadening due to the pure dephasing rate $\gamma_{d,2}$.

\begin{figure*}[htbp]
\centering
\includegraphics[width=2.2in, height=2.11in]{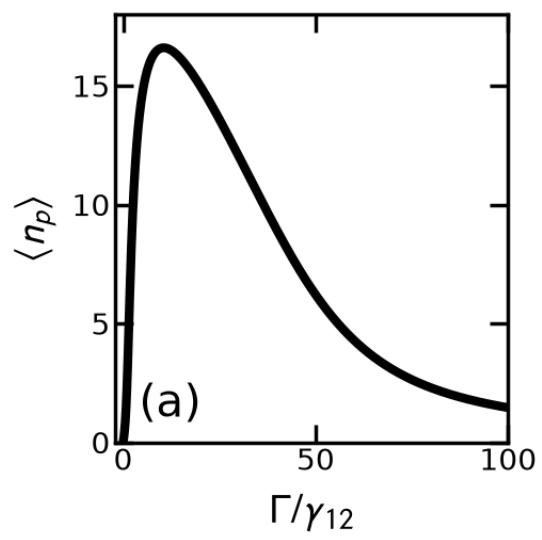} 
\includegraphics[width=2.2in, height=2.11in]{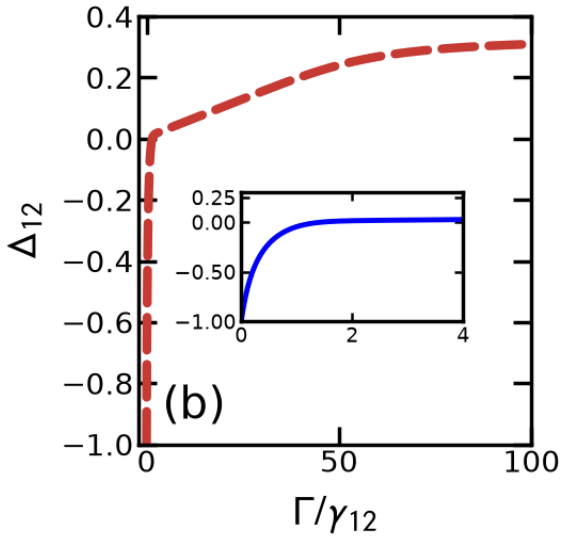} 
\includegraphics[width=2.2in, height=2.11in]{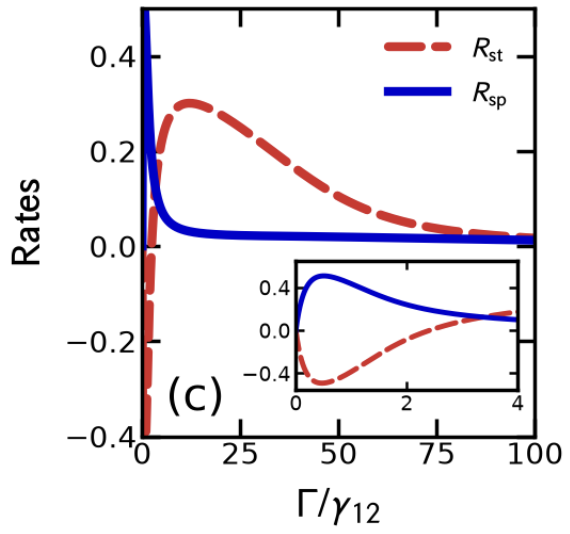} 
\captionsetup{format=plain, margin=1em, justification=raggedright, singlelinecheck=false}
\caption{(Color online) Steady-state cavity observables calculated up to $\Gamma/\gamma_{12} = 100$ using a basis truncation size of $\mathcal{N}=40$. (a)~Mean cavity photon number $\langle n_p \rangle$. (b)~Population inversion $\Delta_{12}$ on the $|2\rangle \to |1\rangle$ transition; the inset shows the behavior across the low-pump region ($\Gamma/\gamma_{12} \le 4$). (c)~Radiative decay rates comparing the net stimulated emission rate $R_{\text{st}}$ (dashed red curve) and the spontaneous emission rate $R_{\text{sp}}$ (solid blue curve); the inset shows a magnified view of the low-pump region to highlight the rate-crossing point. Parameters used in units of $\gamma_{12}$ are $\gamma_{13} = 0.00$, $\gamma_{23} = 0.50$, $g = 1.00$, and $\kappa = 0.01$, corresponding to a cooperativity of $\mathcal{C}=100$.}
\label{Fig2}
\end{figure*}

The numerical results over the full parameter range are shown in Fig.~\ref{Fig2}. In the wide-range sweep presented in Fig.~\ref{Fig2}(a), we find that the average intracavity photon number $\langle n_p \rangle$ reaches a maximum of $\langle n_p \rangle \approx 16.5$ at a normalized pumping rate of $\Gamma/\gamma_{12} \approx 10$. Because a large fraction of the spontaneous emission is coupled into the single resonant cavity mode due to the Purcell effect, the intensity scales up continuously from zero. Consequently, we notice that the system does not exhibit the sharp threshold knee characteristic of conventional multi-atom ensemble lasers \cite{Gerasimov2021}. Additionally, our results in Fig.~\ref{Fig2}(b) demonstrate that a net positive population inversion ($\Delta_{12} > 0$) is established across the gain transition. We observe that this inversion crosses the transparency boundary ($\Delta_{12} = 0$) at a pumping rate of $\Gamma/\gamma_{12} \approx 1.25$.

By evaluating the radiative rates defined in Eqs.~\eqref{rst_trace_eq} and \eqref{rsp_trace_eq}, we identify an operational threshold for the single-emitter system. As shown in the Fig.~\ref{Fig2}(c), at low pumping rates where $\Gamma/\gamma_{12} < 2.45$ (see the inset plot for clarity), the net stimulated emission rate $R_{\text{st}}$ remains below the spontaneous emission rate $R_{\text{sp}}$. In this sub-threshold regime, although the system achieves optical transparency past $\Gamma/\gamma_{12} \approx 1.25$ (where absorption exactly balances stimulated emission) stimulated emission does not dominate the cavity dynamics, and the field is primarily driven by spontaneous fluctuations. However, at $\Gamma/\gamma_{12} \approx 2.45$, we observe a crossing point where the net stimulated emission rate equals the spontaneous emission rate ($\Delta R \equiv R_{\text{st}} - R_{\text{sp}} = 0$). Past this point, stimulated processes become the dominant mechanism driving the cavity field dynamics, marking the structural transition into the lasing regime. We now turn to a detailed examination of the multi-photon subspaces and how we validate the numerical stability of our results in our three operating regimes of interest.

\begin{figure*}
\centering
\includegraphics[width=0.48\textwidth]{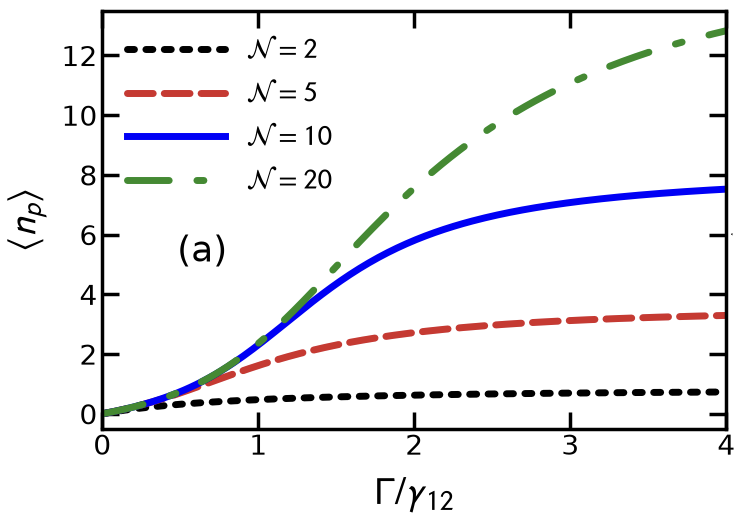}\hfill
\includegraphics[width=0.48\textwidth]{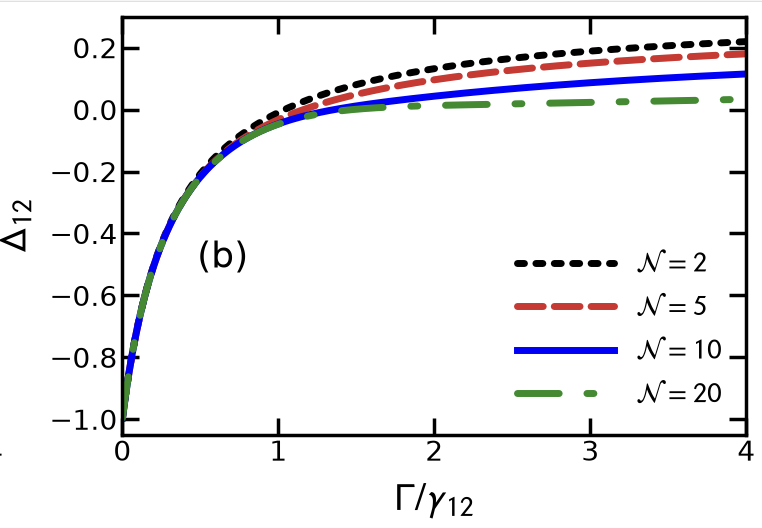}
\vspace{4mm} 
\includegraphics[width=0.274\textwidth]{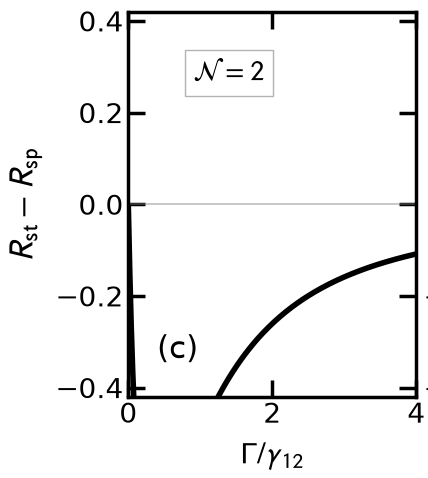}\hfill
\includegraphics[width=0.24\textwidth]{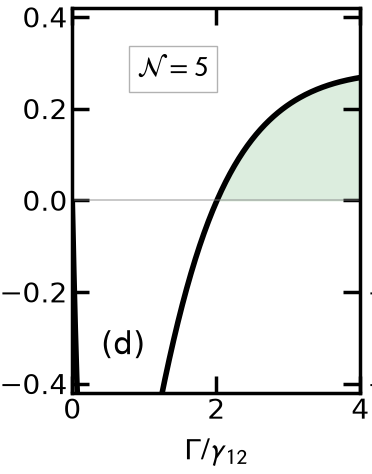}\hfill
\includegraphics[width=0.24\textwidth]{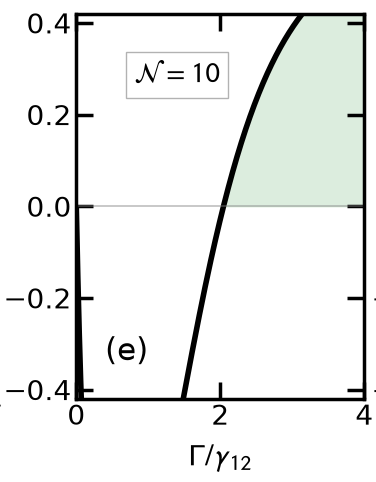}\hfill
\includegraphics[width=0.24\textwidth]{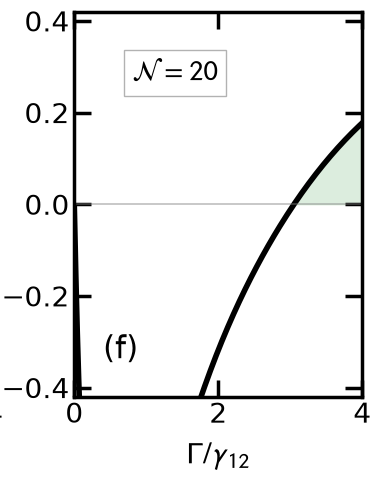}
\captionsetup{format=plain, margin=1em, justification=raggedright, singlelinecheck=false}
\caption{(Color online) Cavity observables calculated as a function of the normalized pump rate $\Gamma/\gamma_{12}$ for different basis truncation sizes $\mathcal{N} = 2, 5, 10,$ and $20$. (a)~Mean cavity photon number $\langle n_p \rangle$. (b)~Population inversion $\Delta_{12}$ on the $|2\rangle \to |1\rangle$ transition. (c)--(f)~Net emission rate balance $\Delta R = R_{\text{st}} - R_{\text{sp}}$ calculated for truncation cutoffs of (c)~$\mathcal{N}=2$, (d)~$\mathcal{N}=5$, (e)~$\mathcal{N}=10$, and (f)~$\mathcal{N}=20$. Shaded regions indicate the pumping domain where the net stimulated rate exceeds the spontaneous emission rate ($\Delta R > 0$). The parameters used in this simulation in the units of $\gamma_{12}$ are given by $\gamma_{13} = 0.00$, $\gamma_{23} = 0.50$, $g = 1.00$, and $\kappa = 0.01$.}
\label{Fig3_convergence}
\end{figure*}

\subsection{\label{subsec:multiphoton_convergence} Multi-Photon Subspaces and Truncation Validity}
We begin this subsection by noticing that rather than treating the computational basis size strictly as a parameter for algorithmic convergence, systematically expanding the truncation limit across $\mathcal{N} = 2, 5, 10,$ and $20$ provides a physical metric for evaluating multi-photon correlations. We thus find from this comparative analysis how artificially low truncation limits alter the calculated light-matter properties, which are only accurately captured when the basis is sufficiently large to include high-dimensional Fock states. As shown in Figs.~\ref{Fig3_convergence}(a) and 3(b), we find by restricting the system to a low-dimensional subspace where $\mathcal{N} = 2$ limits the mean cavity photon number to a maximum of $\langle n_p \rangle \approx 0.77$. This restriction also forces the population inversion $\Delta_{12}$ into an artificial plateau. Physically, this behavior occurs because a severely truncated basis excludes multi-photon state mixing, failing to account for higher-order quantum fluctuations. Expanding the basis cutoff to higher dimensions shows that the atomic inversion does not saturate early, allowing the multi-photon field to scale continuously toward its true steady-state maximum of $\langle n_p \rangle \approx 16.5$.

Additionally, we observe that such truncation restrictions become further apparent in the net radiative rate balances, $\Delta R = R_{\text{st}} - R_{\text{sp}}$, calculated for individual basis sizes in Figs.~\ref{Fig3_convergence}(c)–(f). For the lowest truncation limit of $\mathcal{N} = 2$ (Fig.~\ref{Fig3_convergence}(c)), the rate balance remains negative across the entire range of pumping rates. This result incorrectly indicates that stimulated emission cannot overcome the spontaneous emission noise floor within this restricted subspace. By expanding the basis dimension to include higher multi-photon configurations ($\mathcal{N} = 5 \to 20$), panels \ref{Fig3_convergence}(d)–(f) illustrate the development of a positive gain window, indicated by the shaded green regions. This progression proves that a high-dimensional calculation is required to accurately resolve the stimulated amplification window. Tracking this boundary establishes the minimum basis sizes needed to incorporate high-order Fock state populations and identify the true threshold crossings of the single-emitter lasers.

\subsection{Intermediate Intensity Domain and Non-Classical Statistics}
Keeping in view of the results presented above, we now focus on the intermediate driving regime ($2 \le \Gamma/\gamma_{12} \le 25$) to investigate the development of macroscopically populated cavity fields and the corresponding photon statistics in our lasing regime. To this end, we solve the time-dependent master equation for the total density matrix $\rho(t)$ and trace out the atomic degrees of freedom to isolate the reduced cavity field density operator:
\begin{equation}
\hat{\rho}_{\text{field}}(t) = \text{Tr}_{\text{atom}}\{\hat{\rho}(t)\} = \sum_{i=1}^{3} \langle i|\hat{\rho}(t)|i\rangle. \label{eq:12}
\end{equation}
Using the field matrix elements $\rho_{n,m}(t) \equiv \langle n|\hat{\rho}_{\text{field}}(t)|m\rangle$, we evaluate the phase-space characteristics of the cavity field through the time-resolved Wigner quasiprobability distribution function $W(\alpha)$ \cite{schleich2015quantum} across the complex displacement coordinates $\alpha = x + ip$. Utilizing the canonical Laguerre polynomial expansion, the distribution is computed as:
\begin{align}
W(\alpha) = \frac{2}{\pi} \sum_{n=0}^{\mathcal{N}-1} \sum_{m=0}^{\mathcal{N}-1} \rho_{n,m} (-1)^n \sqrt{\frac{n!}{m!}} (2\alpha^*)^{m-n} L_n^{m-n}(4|\alpha|^2), \label{eq:13}
\end{align}
where $L_n^{m-n}(x)$ denotes the associated Laguerre polynomial. Concurrently, the diagonal components determine the time-dependent photon-number probability distribution across multi-photon configurations: $P(n, t) = \rho_{n,n}(t) = \langle n|\hat{\rho}_{\text{field}}(t)|n\rangle$.

Furthermore, to quantify the steady-state intensity fluctuations and noise variance profiles across the full pumping spectrum, we calculate the zero-delay second-order coherence function $g^{(2)}(0)$ \cite{fox2006quantum} and the Mandel parameter $Q$ \cite{mandel1979sub} using the steady-state density operator $\rho_{ss}$ as:
\begin{equation}
g^{(2)}(0) = \frac{\text{Tr}\{\hat{a}^\dagger \hat{a}^\dagger \hat{a} \hat{a} \hat{\rho}_{ss}\}}{(\text{Tr}\{\hat{a}^\dagger \hat{a} \hat{\rho}_{ss}\})^2} = \frac{\sum_{n=0}^{\mathcal{N}-1} n(n-1)P(n)}{\left(\sum_{n=0}^{\mathcal{N}-1} nP(n)\right)^2}, \label{eq:14}
\end{equation}
\begin{equation}
Q = \frac{\langle(\hat{a}^\dagger \hat{a})^2\rangle - \langle \hat{a}^\dagger \hat{a} \rangle^2 - \langle \hat{a}^\dagger \hat{a} \rangle}{\langle \hat{a}^\dagger \hat{a} \rangle} = \langle n_p \rangle \left[g^{(2)}(0) - 1\right]. \label{eq:15}
\end{equation}

\begin{figure*}
\centering
\includegraphics[width=0.32\textwidth]{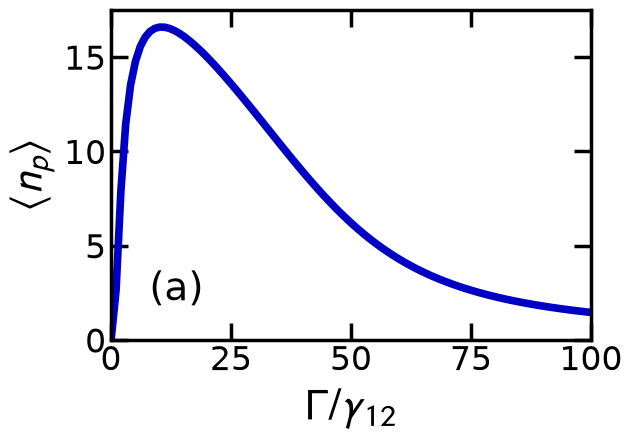}\hfill
\includegraphics[width=0.34\textwidth]{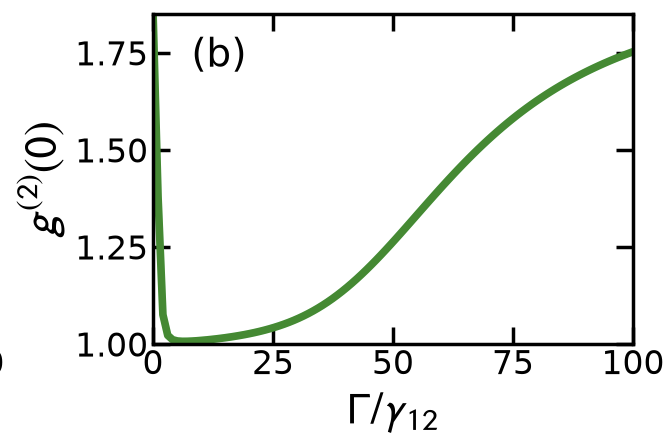}\hfill
\includegraphics[width=0.32\textwidth]{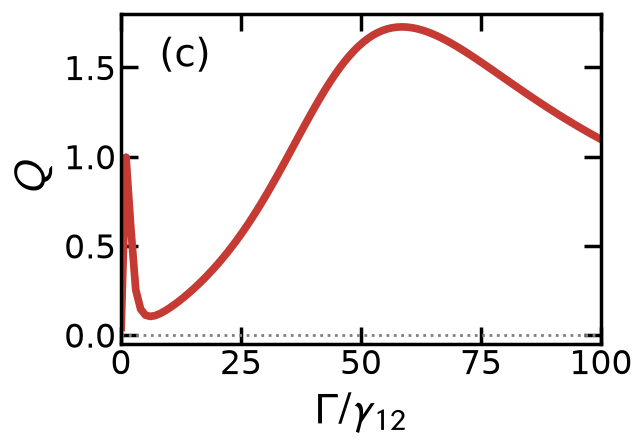}
\vspace{5mm} 
\includegraphics[width=0.232\textwidth]{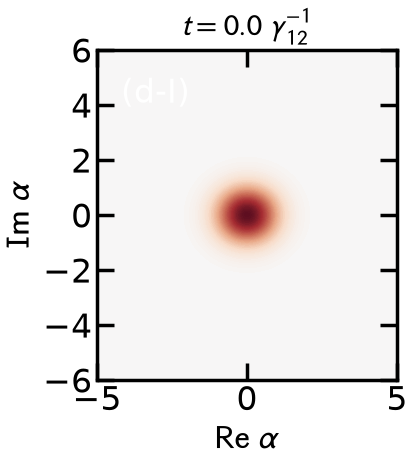}\hfill
\includegraphics[width=0.187\textwidth]{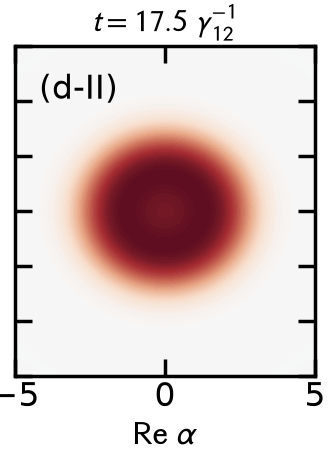}\hfill
\includegraphics[width=0.187\textwidth]{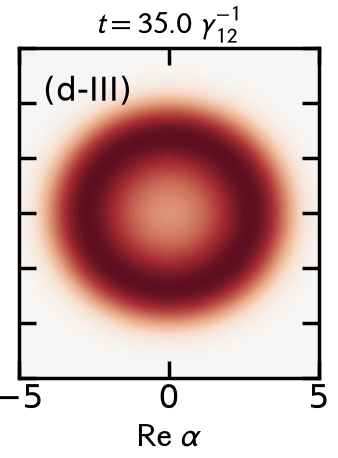}\hfill
\includegraphics[width=0.192\textwidth]{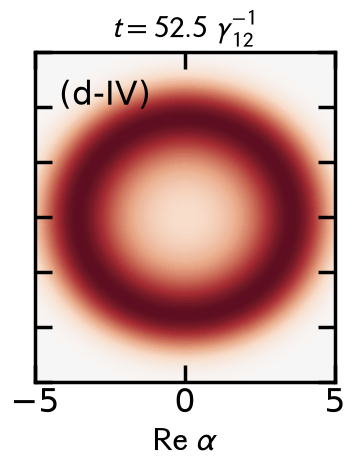}\hfill
\includegraphics[width=0.192\textwidth]{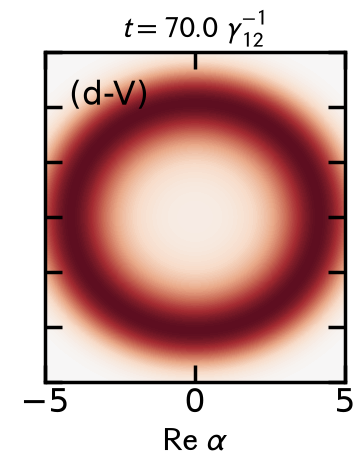}
\vspace{2mm} 
\includegraphics[width=0.234\textwidth]{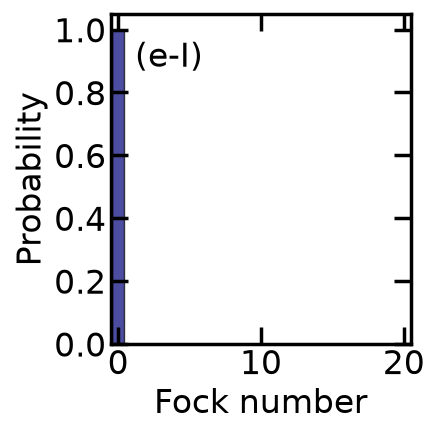}\hfill
\includegraphics[width=0.185\textwidth]{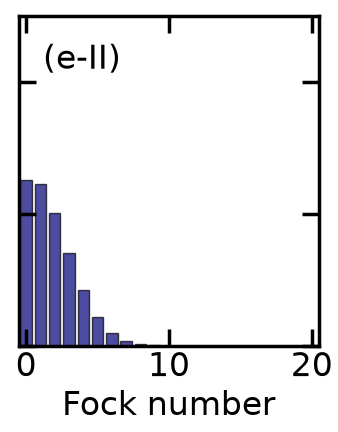}\hfill
\includegraphics[width=0.185\textwidth]{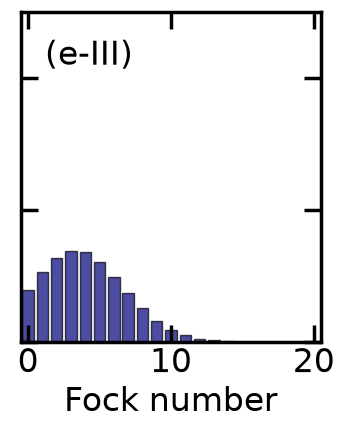}\hfill
\includegraphics[width=0.185\textwidth]{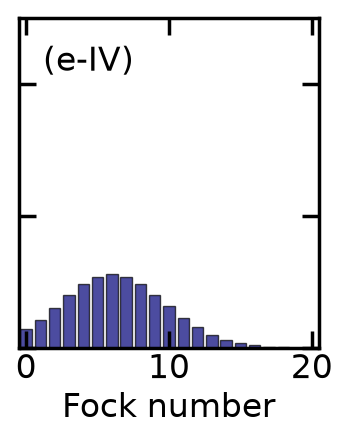}\hfill
\includegraphics[width=0.185\textwidth]{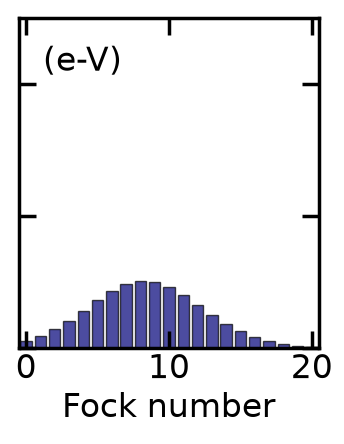}
\captionsetup{format=plain, margin=1em, justification=raggedright, singlelinecheck=false}
\caption{(Color online) Steady-state results and transient field buildup calculated under a convergent computational basis choice of $\mathcal{N}=60$. (a)~Mean cavity photon population $\langle n_p \rangle$, (b)~Zero-delay second-order coherence function $g^{(2)}(0)$, and (c)~the Mandel $Q$-parameter. All curves plotted as a function of normalized pumping rate $\Gamma/\gamma_{12}$. (d-I)--(d-V)~Time-resolved phase-space snapshots tracking the radial expansion of the Wigner distribution function $W(\alpha)$ from vacuum at intervals $t = \{0.0, 17.5, 35.0, 52.5, 70.0\}\,\gamma_{12}^{-1}$. (e-I)--(e-V)~Transient evolution of the corresponding photon-number probability distribution $P(n)$. The parameters used (in units of $\gamma_{12}$) in these plots are $\gamma_{13} = 0.00$, $\gamma_{23} = 0.50$, $g = 1.00$, $\kappa = 0.01$.}
\label{Fig4_coherence_matrix}
\end{figure*}

We plot the steady-state mean cavity photon population $\langle n_p \rangle$ as a function of the normalized pump rate in Fig.~\ref{Fig4_coherence_matrix}(a). We notice the intensity reaches a peak value of $\langle n_p \rangle \approx 16.5$ before undergoing high-pump suppression. In contrast to macroscopic ensemble lasers, the steady-state results presented in Figs.~\ref{Fig4_coherence_matrix}(b) and \ref{Fig4_coherence_matrix}(c) demonstrate that the emitted field remains strictly super-Poissonian ($g^{(2)}(0) > 1$ and $Q > 0$) across the entire range of pumping rates considered. Within the optimal lasing window, we notice that increasing the pumping rate causes the amplitude noise to diminish considerably. In particular, when we plot the Mandel parameter in Fig.~\ref{Fig4_coherence_matrix}(c) we find that it reaches a sharp minimum value of of $Q \approx 0.1$ near $\Gamma/\gamma_{12} \approx 6$, indicating an amplitude-stabilized multi-photon state. We attribute this behavior to emerge from the persistent incoherent pumping fluctuations and spontaneous fluctuations inherent to a single emitter, such that the classical shot-noise limit forms a strict physical lower bound ($Q = 0, g^{(2)}(0) = 1$) here. 

It is worthwhile to note from Fig.~\ref{Fig4_coherence_matrix}(a) that when the driving strength enters the over-pumping regime ($\Gamma/\gamma_{12} > 25$), we observe that the mean cavity photon number monotonically decays toward the vacuum state. We argue that this decay occurs because intense incoherent pumping populates the highest excited electronic level $|3\rangle$ at a rate faster than other decaying transitions can deplete it, trapping the atomic population in this upper state ($\langle \sigma_{33} \rangle \to 1$). As a result, the intermediate relaxation governed by $\gamma_{23}$ begins to act as a excitation bottleneck that restricts the population of the active lasing transition level $|2\rangle$. Under these driving conditions, we notice that the cavity field coherence degrades rapidly, causing $g^{(2)}(0)$ to rise toward $1.75$ and resulting in a pronounced increase in the Mandel parameter as well, which indicates a clear transition of the emitted light to follow a chaotic thermal statistics.

We plot the transient evolution of the cavity field in Figs.~\ref{Fig4_coherence_matrix}(d-I)--\ref{Fig4_coherence_matrix}(d-V) and Figs.~\ref{Fig4_coherence_matrix}(e-I)--\ref{Fig4_coherence_matrix}(e-V) for a fixed $\Gamma = 10\gamma_{12}$. Initialized in the vacuum state at $t = 0$ (panels (d-I), (e-I)), we find that the quasiprobability density expands radially outward over time (panels (d-II)--(d-V)), forming a local minimum at $\alpha = 0$ and stabilizing into a rotationally symmetric ring. Because our model describes a single three-level emitter rather than a collective ensemble, we note that the system lacks a macroscopic symmetry-breaking mechanism to stabilize a well-defined phase. Instead, incoherent pump fluctuations distribute the phase uniformly along the perimeter, yielding a phase-insensitive ring characteristic of a single-atom laser \cite{mikhalychev2022phase}, which we see is consistent with the broadening of the Fock-state populations $P(n)$ in Figs.~\ref{Fig4_coherence_matrix}(e-I)--(e-V). While the radial displacement confirms a stable multi-photon amplitude, we find that the lack of phase-locking causes the first-order correlation function $\langle \hat{a}^\dagger(t+\tau)\hat{a}(t)\rangle$ to decay exponentially. We expect this phase diffusion to manifest as line-broadening in the cavity emission spectrum which could be resolved through asymmetric Mach-Zehnder interferometry or homodyne detection.

\subsection{\label{subsec:semiclassical_regime} Semiclassical Regime}

We now turn our attention to establish an analytical connection with macroscopic laser models. To this end, we implement a semiclassical treatment which relies on a mean-field approximation \cite{wang2022coherent} that factorizes multi-operator quantum correlations into independent expectation product pairs, enforcing the following decoupling rules:
\begin{align}
    \langle \hat{a}^\dagger \hat{a} \hat{\sigma}_{ii} \rangle &\approx \langle \hat{a}^\dagger \hat{a} \rangle \langle \hat{\sigma}_{ii} \rangle \equiv \langle n_p \rangle_{\text{sc}} p_i, \label{fact_rule_1} \\
    \langle \hat{a}^\dagger \hat{\sigma}_{ij} \rangle &\approx \langle \hat{a}^\dagger \rangle \langle \hat{\sigma}_{ij} \rangle = 0, \label{fact_rule_2}
\end{align}
where $p_i \equiv \langle \sigma_{ii} \rangle$ represents the population probability of level $|i\rangle$, and $\langle n_p \rangle_{\text{sc}}$ denotes the semiclassical photon intensity. By applying these factorization properties to the full master equation, we eliminate the spontaneous emission noise coupling terms ($G_n \langle \sigma_{22} \rangle$) and multi-photon quantum noise channels. This reduces our open quantum system to a closed set of coupled non-linear differential rate equations governing the mean macro-scale parameters as:
\begin{align}
    \frac{d\langle n_p \rangle_{\text{sc}}}{dt} &= \frac{4g^2}{\gamma_{12} + \Gamma} \Delta_{12\text{(sc)}} \langle n_p \rangle_{\text{sc}} - 2\kappa \langle n_p \rangle_{\text{sc}}, \label{sc_photon_eq} \\
    \frac{d p_3}{dt} &= \Gamma p_1 - \gamma_{23} p_3, \label{sc_level3_eq} \\
    \frac{d p_2}{dt} &= \gamma_{23} p_3 - \gamma_{12} p_2 - \frac{4g^2}{\gamma_{12} + \Gamma} \Delta_{12\text{(sc)}} \langle n_p \rangle_{\text{sc}}, \label{sc_level2_eq}
\end{align}
with a condition that $p_1 + p_2 + p_3 = 1$. Not that here $\Delta_{12\text{(sc)}} \equiv p_2 - p_1$ defines the semiclassical population inversion.

\begin{figure*}[htbp]
\centering
\includegraphics[width=0.32\textwidth]{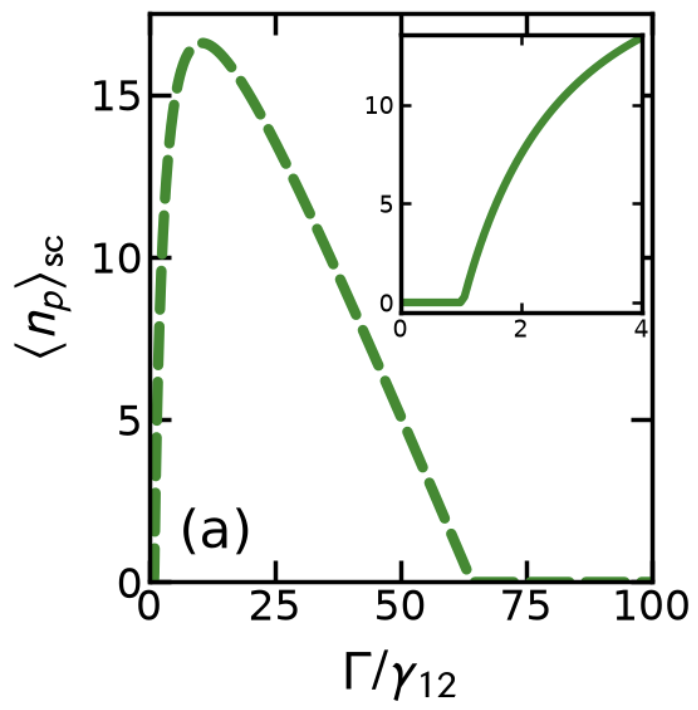}\hfill
\includegraphics[width=0.32\textwidth]{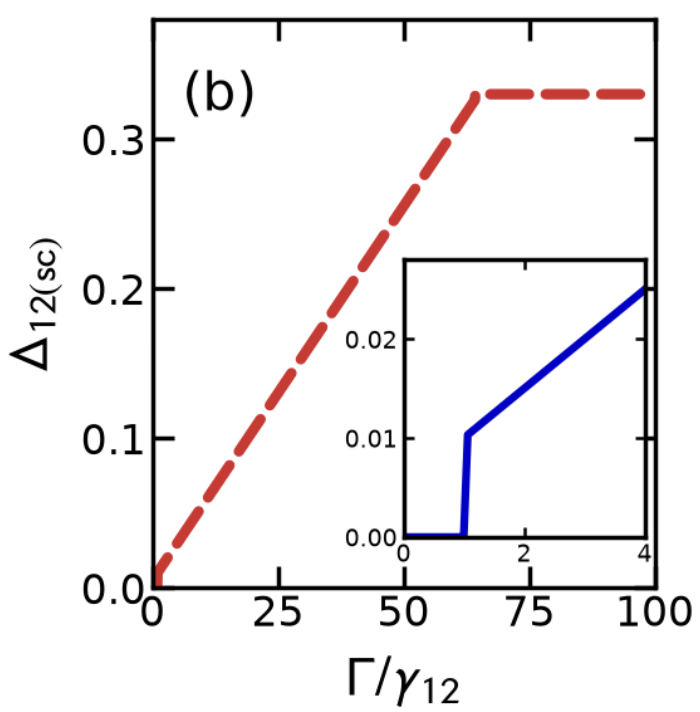}\hfill
\includegraphics[width=0.32\textwidth]{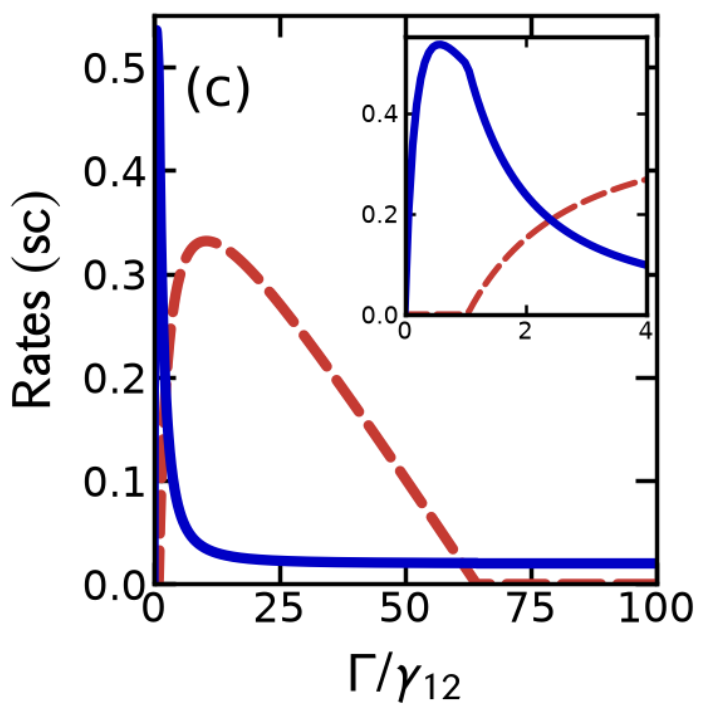}
\captionsetup{format=plain, margin=1em, justification=raggedright, singlelinecheck=false}
\caption{(Color online) Semiclassical steady-state laser profiles calculated over a wide pumping range up to $\Gamma/\gamma_{12} = 100$ using a factorized mean-field approximation. (a) Mean photon number $\langle n_p \rangle_{\text{sc}}$, which peaks near $\Gamma/\gamma_{12} \approx 11$ before dropping to zero past $\Gamma/\gamma_{12} \approx 65$. (b) Atomic population inversion $\Delta_{12\text{(sc)}}$. (c) Semiclassical rates comparing the net stimulated process (dashed red line) with the spontaneous emission process (solid blue line). The insets provide a magnified view of the low-pumping region ($\Gamma/\gamma_{12} \le 4$) to clearly resolve the laser threshold behavior. The rest of the parameters are the same as used in Fig.~\ref{Fig4_coherence_matrix}.}
\label{Fig5_semiclassical_matrix}
\end{figure*}

Next, to find the analytic conditions required for the onset of stimulated amplification (modified by pure electronic dephasing), we set the time derivatives to zero. We thus find
\begin{equation}
    \Delta_{12\text{(sc)}}^{\text{threshold}} = \frac{\kappa (\gamma_{12} + \Gamma + 2\gamma_{d,2})}{2g^2} = \frac{1}{2\mathcal{C}} \left(1 + \frac{\Gamma + 2\gamma_{d,2}}{\gamma_{12}}\right),
\label{Eq18_threshold}
\end{equation}
where, as before, $\mathcal{C} \equiv g^2 / (2\kappa \gamma_{12})$ represents the cooperativity parameter \cite{kimble1998strong, nori2025cavity}. We find that pure electronic dephasing reduces the cavity coherence, which shifts the laser threshold to higher pump rates and promoting the onset of self-quenching at high pump powers. By setting the time derivatives equal to zero, we obtain the analytical steady-state solutions for the lasing regime
\begin{align}
    p_1 &= \frac{1 - \Delta_{12\text{(sc)}}^{\text{threshold}}}{2 + \frac{\Gamma}{\gamma_{23}}} = \frac{1 - \frac{1}{2\mathcal{C}}\left(1 + \frac{\Gamma}{\gamma_{12}}\right)}{2 + \frac{\Gamma}{\gamma_{23}}}, \label{p1_lasing_sol} \\
    p_2 &= p_1 + \Delta_{12\text{(sc)}}^{\text{threshold}} = p_1 + \frac{1}{2\mathcal{C}} \left(1 + \frac{\Gamma}{\gamma_{12}}\right), \label{p2_lasing_sol} \\
    p_3 &= \frac{\Gamma}{\gamma_{23}} p_1. \label{p3_lasing_sol}
\end{align}
Substituting these steady-state population expressions back into the cavity field equation (Eq.~\eqref{sc_photon_eq}) yields the explicit algebraic equation for the mean field intensity profile along the lasing regime:
\begin{align}
    &\langle n_p \rangle_{\text{sc}} = \frac{\Gamma p_1 - \gamma_{12} p_2}{2\kappa}\nonumber\\
    &= \frac{\Gamma - \gamma_{12} - \frac{1}{2\mathscr{C}}\left(1 + \frac{\Gamma}{\gamma_{12}}\right)\left(\Gamma + \gamma_{12} + \frac{\Gamma \gamma_{12}}{\gamma_{23}}\right)}{2\kappa \left(2 + \frac{\Gamma}{\gamma_{23}}\right)}.\label{n_sc_full_sol}
\end{align}
We plot the steady-state behavior of the semiclassical model in Fig.~\ref{Fig5_semiclassical_matrix}. As shown in Fig.~\ref{Fig5_semiclassical_matrix}(a) and its inset, the cavity remains empty until the pump exceeds $\Gamma/\gamma_{12} \approx 1.05$, where the photon population turns on sharply. Above this threshold, the population inversion $\Delta_{12\text{(sc)}}$ is pinned to the analytical limits of Eq.~\eqref{Eq18_threshold}, rising linearly as seen in Fig.~\ref{Fig5_semiclassical_matrix}(b).

We find that neglecting multi-operator quantum correlations (due to the application of the mean field approximation) overestimates the peak cavity field ($\langle n_p \rangle_{\text{sc}} \approx 18.2$ versus the quantum value $\langle n_p \rangle_{\text{q}} \approx 16.5$). Omitting these fluctuations removes a critical spontaneous noise drag from the feedback loop, artificially overpredicting the net stimulated gain by more than $10\%$. In a full quantum treatment, multi-photon fluctuations stabilize the inversion $\Delta_{12}$ closer to zero over a wide pumping range. Accounting for this radiative drag alongside pure dephasing is therefore essential to correctly map such a lasing device opertaional boundaries.

We extend our tracking up to $\Gamma/\gamma_{12} = 100$ to evaluate the high-pump self-quenching regime. Past $\Gamma/\gamma_{12} \approx 15$, increasing the drive suppresses the photon number until it collapses to zero at $\Gamma/\gamma_{12} \approx 65$. This behavior is driven by upper-level population trapping, which causes the inversion in panel (b) to flatline and the stimulated rate (dashed red line) in panel (c) to vanish. Finally, we emphasize that these plots highlight why decoupled semiclassical rate approximations cannot adequately describe single-emitter cavity QED lasing behaviors.

\subsection{\label{subsec:feasibility} Experimental Feasibility}
To discuss the experimental feasibility of our theoretical study, we compare our dimensionless parameter choices onto a realistic solid-state cavity QED platform. For a self-assembled $\text{InGaAs/GaAs}$ quantum dot \cite{opticaq2024dot} embedded in a high-finesse micropillar cavity at cryogenic temperatures ($T \approx 4\text{ K}$), the spontaneous decay rate is typically $\gamma_{12} \approx 1\text{ GHz}$, corresponding to an exciton lifetime of $\tau \approx 1\text{ ns}$ \cite{senellart2017high}. Within this setting, our strong-coupling parameters $g \approx 1\text{ GHz}$ and $\kappa \approx 10\text{ MHz}$ place the laser threshold ($\Gamma/\gamma_{12} \approx 2.45$) at a physical pumping rate of \cite{reitzenstein2010quantum} $\Gamma \approx 2.45\text{ GHz}$. Assuming a standard near-infrared photon energy of $E_{\text{photon}} \approx 1.3\text{ eV}$ and a non-resonant absorption efficiency of $0.2\%$ into the quantum dot inside the cryostat, this threshold corresponds to an incident optical pump power of approximately $1.6\ \mu\text{W}$. This microwatt threshold lies well within modern experimental capabilities \cite{reitzenstein2006lasing, reitzenstein2008single}. Consequently, our system offers a viable platform for high-resolution photon-correlation \cite{ulrich2007photon} and homodyne detection measurements to observe the predicted self-quenching behavior and photon statistics.

\section{\label{sec:IV} Summary and Conclusions}
In summary, we have analyzed the steady-state and transient dynamics of a three-level single-emitter microcavity laser by solving the full open-system Lindblad master equation. By systematically checking our simulations across expanded basis sizes up to $\mathcal{N} = 60$, we show that verifying convergence against truncation limits is essential to avoid numerical truncation artifacts and extract true physical observables. Using this convergent framework, we mapped out the operational thresholds and noise characteristics of individual single-atom lasing sources, providing concrete parameter baselines to optimize cryogenic solid-state systems like quantum dots and color centers.

In particular we conclude that the single-emitter cavity QED lasers operates under a distinct balance of strong-coupling regime and quantum fluctuations. Due to Purcell-enhanced spontaneous emission into the resonant cavity mode, the intracavity field grows continuously without the sharp intensity step typical of macroscopic ensemble lasers. However, by tracking individual radiative decay channels, we isolate a clear threshold at a normalized pumping rate of $\Gamma/\gamma_{12} \approx 2.45$, where the net stimulated emission rate overcomes the spontaneous noise floor. In phase space, the transient Wigner distribution expands into a rotationally symmetric ring. This symmetry confirms that without a macroscopic atomic ensemble to break the $U(1)$ phase symmetry, random phase kicks from the incoherent pump cause complete phase diffusion.

Finally, our steady-state noise calculations showed that the field remains strictly super-Poissonian ($Q > 0, g^{(2)}(0) > 1$) across the entire driving range, as discrete incoherent pumping fluctuations prevent sub-Poissonian squeezing. At high driving rates up to $\Gamma/\gamma_{12} = 100$, the system enters a self-quenching regime where the cavity field collapses back to zero due to population trapping in the upper excited state. Comparing the exact master equation results with a factorized mean-field approximation reveals that semiclassical models successfully predict the threshold turn-on and self-quenching points, but omit a critical quantum drag force, causing them to overestimate the peak field intensity ($\langle n_p \rangle_{\text{sc}} \approx 18.2$ versus $\langle n_p \rangle_{\text{q}} \approx 16.5$). These results highlight why full quantum treatments are necessary to evaluate the precise performance profiles of single-emitter lasing platforms.

\acknowledgments
This work is dedicated to the memory of Prof. Perry Rice (1959–2026), emeritus professor of physics at Miami University of Ohio. The authors thank Quincy Webb for performing some of the preliminary calculations. IMM acknowledges support from Miami University through the Quantum Physics Collaborative and the Ohio Institute for Quantum Computing.

\bibliographystyle{ieeetr}
\bibliography{sample}
\end{document}